\documentclass[showpacs,pra,aps,twocolumn,float,floatfix]{revtex4}
\usepackage{graphicx,epsfig}
\usepackage{times}
\usepackage{amsmath,amssymb}

\DeclareMathOperator{\re}{Re}
\DeclareMathOperator{\im}{Im}

\newlength\figurewidth
\newcommand{\rem}[1]{}

\newcommand{\realb}[1]{\text{Re}[#1]}

\newcommand{\imagc}[1]{\text{Im}\,#1}
\newcommand{\realc}[1]{\text{Re}\,#1}

\begin{document}

\title{Lifetime statistics in chaotic dielectric microresonators}
\author{Henning Schomerus}
\affiliation{Department of Physics, Lancaster University, Lancaster LA1 4YB, United Kingdom}
\author{Jan Wiersig}
  \affiliation{Institut f{\"u}r Theoretische Physik, Universit{\"a}t Magdeburg,
  Postfach 4120, D-39016 Magdeburg, Germany}
\author{J{\"o}rg Main}
\affiliation{1. Institut f{\"u}r Theoretische Physik, Universit{\"a}t
  Stuttgart, 70550 Stuttgart, Germany}
\date{December 2008}

\begin{abstract}
We discuss the statistical properties of lifetimes of electromagnetic
eigenmodes in dielectric microresonators with fully chaotic ray dynamics.
Using the example of a resonator of stadium geometry, we find that a recently
proposed random-matrix model very well describes the lifetime statistics of
long-lived resonances, provided that two effective parameters are
appropriately renormalized. This renormalization is linked to the formation
of anomalously short-lived resonances, a mechanism also known from the
fractal Weyl law and the resonance trapping phenomenon.
\end{abstract}
\maketitle

\section{Introduction}

Over the past decade optical microresonators have attracted
considerable attention in various fields of
physics~\cite{Vahala03}. Microresonators are basic components for
single-photon emitters~\cite{Michler2000}, ultralow threshold
lasers~\cite{Park04,Ulrich06}, and solid-state cavity quantum
electrodynamics~\cite{RSL04,PSMLHGB05,Schwab06}, to name just a
few of the possible applications. Deformed dielectric microdisk
resonators are of particular interest since they allow for
directional light emission, an effect that can be understood and
optimized by exploring fundamental principles of ray-wave
correspondence~\cite{ND97,GCNNSFSC98,CTSCKJ03,WH06,WH08,SCL08}.
This directionality can be employed to design microlasers which
make best use of the total output intensity. Whether a carefully
designed mode admits substantial lasing depends on how its
lifetime compares to the lifetimes of other modes in the
amplification window. This information is encoded in the imaginary
parts of the resonance frequencies. However, the associated
spectral statistics have been studied only for a few special
geometries, such as the circular disk without and with surface
roughness~\cite{RLRPK07,SJNS00}, while other investigations
focussed on disordered media~\cite{random1,random2} and resonators
with small
openings~\cite{fyodorov:sommers,ballistic1,sommers:fyodorov,ballistic2,ballistic3,ballistic4,KHMS08}.

In order to address the spectral properties of dielectric microdisk
resonators from a broader perspective it is desirable to adopt concepts which
have been successfully applied to other classes of open systems. In
particular, the characterization of spectral statistics is often based on two
ingredients: (a) A Weyl law~\cite{weyllaw}, which concerns the average number
of modes within a given frequency interval, and (b) random-matrix theory
(RMT)~\cite{rmt}, which addresses the statistical fluctuations and
correlations of these modes. Traditionally both concepts have been applied to
systems with small openings, for which the Weyl law is identical to that of
the closed counterpart, while RMT can be justified because modes become
strongly mixed on the scale of the typical life time.

A framework how to adapt these concepts to dielectric microdisk resonators,
who are relatively more open systems allowing for refractive escape along the
entire boundary, has only emerged very recently. Due to the short typical
lifetime, strong mode mixing is only observed for a reduced number of modes.
For systems with ballistic escape, where the mode-mixing time scale is given
by the Ehrenfest time
\cite{Ehrenfesttimeorig,Ehrenfesttimedyn,brouwer,whitney} and the support of
the long-living states is the fractal repeller, it has been established that
the number of modes that exceed a certain lifetime  is given by a {\em
fractal} Weyl law
~\cite{fractalweyllaw1,fractalweyllaw2,schomerus:tworzydlo,fractalweyllaw3,fractalweyllaw4}.
Based on a generalization of the repeller for refractive escape, the
significance of the fractal Weyl law to dielectric microdisk systems with
small refractive index has been demonstrated for the stadium geometry in
Ref.~\cite{WiersigMain08}. In the present work we consider the same geometry
and explore the predictive power of RMT, whose extension to dielectric
microresonators was proposed in Ref.~\cite{Keating}. The application of this
extension requires to determine two effective parameters (a time scale for
the internal dynamics and an effective number of modes), which are sensitive
to the formation of short living resonances (a mechanism also at heart of the
fractal Weyl law~\cite{schomerus:tworzydlo}, as well as the resonance
trapping phenomenon \cite{resonancetrapping}). We concentrate on moderately
large values of the refractive index, for which the short-living resonance
states resemble the bouncing-ball modes of the closed stadium billiard
\cite{bouncingballs}, and find that the numerical lifetime statistics of
long-living modes agree very well with the RMT predictions.

The outline of this paper is as follows. The background Section
\ref{sec:2} provides a brief introduction into the spectral
properties of dielectric microdisk resonators and the dynamical
model of Ref.~\cite{Keating}. In Section~\ref{sec:3} we present
numerical results for complex resonance frequencies in
stadium-shaped microresonators and then analyze the statistics of
long-living resonances in two steps. First we show that the mean
properties of the resonance frequencies can be characterized in
terms of the two effective parameters mentioned above. Based on
these parameters, we secondly  show that fluctuations of the life
times are well captured by RMT. The results are summarized in the
concluding Sec.\ \ref{sec:4}.

\section{\label{sec:2}Theoretical background}
\subsection{Dielectric microdisk resonators}

Dielectric microdisk resonators effectively have a 2D geometry, defined by a
region of constant refractive index $n$ which is embedded into vacuum (where
the refractive index is unity). The classical dynamics inside such a
resonator is similar to a billiard, where a point-like particle moves freely
in a two-dimensional domain with elastic reflections at the
boundary~\cite{Berry81}. Depending on the shape of the boundary a billiard
can show a variety of dynamical behaviors ranging from integrable to fully
chaotic~\cite{Robnik83}. Light rays in a microdisk behave similarly as they
are totally reflected at the interfaces as long as the angle of incidence is
larger than the critical angle for total internal reflection. If, however,
the angle of incidence is smaller than the critical angle, light can escape
refractively according to Snell's and Fresnel's laws. Hence, dielectric
microdisk resonators are {\it leaking
billiards}~\cite{ND97,GCNNSFSC98,NKLSW07}.

The leakiness of the resonator limits the photon lifetime, which can be
described by associating complex values to the resonance frequencies
$\omega$. These frequencies can be obtained from the Helmholtz
equation~\cite{Jackson83eng}
\begin{equation}\label{eq:wave} -\nabla^2\psi =
n^2(x,y)\frac{\omega^2}{c^2}\psi \ ,
\end{equation}
where $c$ is the speed of light in vacuum. For a piecewise constant
refractive index, this equation holds for both transverse magnetic (TM) and
transverse electric (TE) polarization. For TM polarization the electric field
is perpendicular to the resonator plane with $E_z =
\realb{\psi(x,y)e^{-i\omega t}}$, and the wave function $\psi$ and its normal
derivative $\partial_\perp\psi$ are continuous across the boundary of the
resonator. For TE polarization, $\psi$ represents the $z$-component  $H_z$ of
the magnetic field, and the functions $\psi$ and $n^{-2}\partial_\perp\psi$
are continuous across the boundary. For plane interfaces, these boundary
conditions yield the Fresnel reflection amplitudes
\begin{subequations}
\label{fresnel}
\begin{eqnarray}
r_{\rm
TM}(p)=\frac{\sqrt{1-p^2}-\sqrt{n^{-2}-p^2}}{\sqrt{1-p^2}+\sqrt{n^{-2}-p^2}}
,
\\
r_{\rm
TE}(p)=\frac{\sqrt{1-p^2}-n\sqrt{1-n^{2}p^2}}{\sqrt{1-p^2}+n\sqrt{1-n^{2}p^2}}
\end{eqnarray}
\end{subequations}
for resonator photons with dimensionless transverse momentum $p=\sin\chi$,
where $\chi$ is the angle of incidence.

At infinity, outgoing wave conditions are imposed which results in
quasi-bound states with  frequencies $\omega$ situated in the lower half of
the complex plane. Whereas the real part is the usual frequency, the
imaginary part is related to the lifetime $\tau=-1/[2\,\imagc{\omega}]$. The
quality factor of a quasi-bound state is defined by $Q =
-\realc{\omega}/[2\,\imagc{\omega}]$.

An alternative method to obtain the quasibound states sets out
with the scattering matrix $S$, which relates the amplitudes of
the incoming wave components  to the amplitudes of the outgoing
wave components  (see, e.g., Refs.\
\cite{fyodorov:sommers,ballistic1,ballistic2,ballistic3,ballistic4}).
At real frequencies, the scattering matrix is unitary. The
condition of a quasibound state corresponds to the situation
where outgoing wave components are admitted in absence of any
incoming radiation. Under these conditions the scattering matrix
diverges. The resonance frequencies therefore correspond to the
poles of the scattering matrix in the complex frequency plane.

\subsection{Dynamical model of microresonators}

Random-matrix theory delivers a statistical description of complex quantum
systems on the basis of just a few system parameters. In particular, a
classification according to fundamental symmetries such as time-reversal
symmetry or spin-rotation symmetry allows to identify ten classes, among
which three are the traditional Wigner-Dyson classes~\cite{rmt}, while a
further seven classes apply to systems with chiral or particle-hole
symmetries~\cite{zirnbauer}. For each symmetry class one can consider two
basis variants of RMT, which either concerns hermitian matrices $H$
(representing the Hamiltonian) or unitary matrices $U$ (representing the
scattering matrix or the time evolution operator). In the absence of
magneto-optical effects photonic systems are represented by the orthogonal
symmetry class, in which the hermitian and unitary matrices are symmetric
under transposition ($H=H^T$, $U=U^T$).

Traditional random-matrix theory does not assume any further
knowledge of the system, which leads to a uniquely defined
invariant probability measure (the Haar measure). Physical
systems, however, often contain additional commonalities.
Dielectric microresonators are a good example, since the escape is
governed by well structured Fresnel laws of reflection and
refraction. These laws not only affect the far-field radiation
characteristics, but also determine the confinement, and therefore
impact on the spectral properties. It is therefore desirable to
incorporate the specific confinement characteristics of dielectric
microresonators into RMT. In order to study spectral statistics,
one furthermore needs to accommodate frequency correlations
associated to the internal dynamics in the resonator. Both
requirements are met by a variant of RMT recently proposed in
Ref.~\cite{Keating}. In this model the scattering matrix takes
the form
\begin{subequations}\label{eq:rmtall}%
\begin{equation}
S=-{\cal R}+{\cal T}U[\exp(-i\omega\tau)-{\cal R}U]^{-1}{\cal T},
\label{eq:rmtall1}
\end{equation}
where the diagonal matrices \begin{equation} {\cal
R}_{lm}=\delta_{lm}r(p_l), \quad {\cal
T}_{lm}=\delta_{lm}\sqrt{1-|r(p_l)|^2} \label{eq:r}
\end{equation}
\end{subequations}
 are determined from the
Fresnel reflection amplitudes (\ref{fresnel}) at discrete impact
parameter $p_l=(2l-M-1)/M$ (with $M$=\,{\rm dim}\,S and $l=1,\ldots,M$)
\cite{remark:correction}. The frequency dependence is
characterized by a time scale $\tau$ which represents the typical
time of flight through the resonator. The $M$-dimensional matrix
$U$ can be interpreted as a unitary time evolution operator which
propagates photons between encounters with the interface.

In the dynamical model (\ref{eq:rmtall}), the poles of the scattering matrix
are determined by the eigenvalue equation
\begin{equation}
{\cal R}U\psi=\exp(-i\theta)\psi,
\label{eq:evals}
\end{equation}
where $\theta=\omega\tau$ is a dimensionless quantity. The
operator on the left is subunitary, and therefore each resonance
frequency in general possesses a negative imaginary part related
to the lifetime of the state. Eigenvalue problems of this type
with various  specifications of ${\cal R}$ have recently attracted
considerable attention
\cite{truncatedrmt1,truncatedrmt2,wei:fyodorov,schomerus:tworzydlo,fractalweyllaw3,fractalweyllaw4,nonnenmacher:schenck}.

In the RMT version of the dynamical model (\ref{eq:rmtall}), $U$
is taken from the circular orthogonal ensembles. The statistical
description is then completed by the specification of the two
parameters $M$ and $\tau$.  Once these parameters are determined,
one can investigate statistical fluctuations. We now implement
this strategy for a specific model system.

\section{\label{sec:3}Life time statistics in the stadium microresonator}

\begin{figure}[t]
\includegraphics[width=0.65\figurewidth]{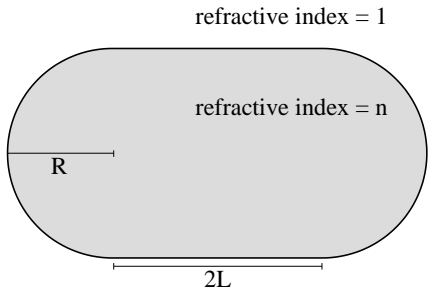}
\caption{Sketch of a stadium-shaped dielectric microresonator with $L=R$.}
\label{fig:sketch}
\end{figure}

A microdisk resonator with stadium-shaped cross-section is illustrated in
Fig.~\ref{fig:sketch}. The region with refractive index $n$ is bounded by two
semicircles with radius $R$ and two straight lines of length $L$. The
corresponding closed cavity, the stadium billiard, is a paradigm for fully
developed chaos~\cite{Bunimovich74b,Bunimovich79}. We consider the case $L =
R$, for which the closed classical billiard has the largest Lyapunov
exponent~\cite{BenS78}.

The open stadium resonator has been studied both theoretically and
experimentally in a number of works, with focus on the spatial mode structure
and ray-wave correspondence~\cite{SHTS06,FHW06,LLZSB07,SH07,SFH07}. In a
first step towards the characterization of the resonance frequency spectrum
of the stadium resonator, the applicability of the Weyl law has been
investigated in Ref.~\cite{WiersigMain08}.

We start our statistical analysis of the spectral fluctuations with
the numerical determination the resonance frequencies. In the next
step, we relate the averages of the real and imaginary parts of
these frequencies to the two parameters $\tau$ and $M$ of the
dynamical model (\ref{eq:rmtall}). This fixes the RMT version of
this model and delivers a prediction for the distribution of life
times, which we compare in the final step to the actual life time
statistics of the stadium resonator.

\subsection{Resonance frequencies}

The resonance states in the stadium billiard can be classified by
their parity with respect to the horizontal and vertical symmetry
lines of the stadium billiard. In the computations this can be
exploited by considering four desymmetrized versions of the
resonator, which differ by the boundary conditions imposed on the
two symmetry lines (Dirichlet or Neumann boundary conditions for
states of odd or even parity with respect to the given symmetry
line, respectively). As explained in detail in Ref.~\cite{WiersigMain08} the
boundary element method~\cite{Wiersig02b} and the harmonic
inversion technique~\cite{Main99} are very efficient tools to compute the
quasibound modes. The
combination of these two methods allows us to gather a large
number of resonant frequencies~$\omega$, as is required for a
reliable statistical analysis. A dimensionless form of these
frequencies can be obtained by considering the scaled frequencies
$\Omega = \omega R/c$. For the present study we collected
resonance frequencies in the interval $66< n \re\Omega < 82.5$ for
both polarizations (TM and TE), all parities, and two different values of the refractive
index, $n=3.3$ (for GaAs) and $n=5$. The latter value we have chosen since a
broad interval of refractive indices allows a more convincing validation of our
theory.

\begin{figure}[t]
\includegraphics[width=\figurewidth]{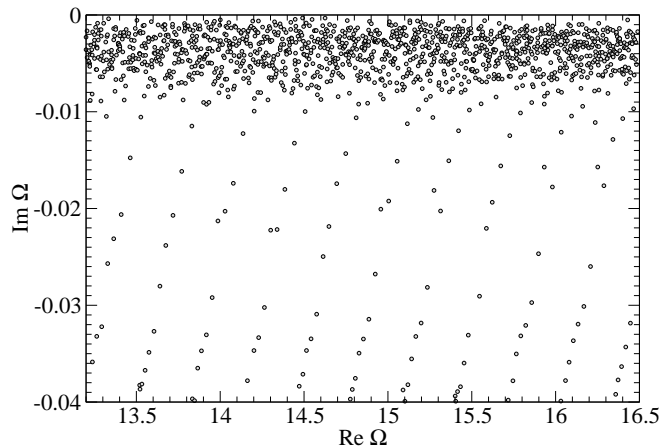}
\caption{Position of dimensionless complex resonance frequencies
with TM polarization for the stadium microresonator with
refractive index $n=5$.} \label{fig:complexplane}
\end{figure}

For illustration, the position of the dimensionless resonance frequencies in
the complex plane for $n=5$ and TM polarization (all parities) is shown in
Fig.\ \ref{fig:complexplane}. The figure shows a clear separation of
short-living resonances which are organized along well-structured bands ($\im
\Omega \lesssim -0.015$), and long living resonances whose position does not
suggest any clear organization principle ($\im \Omega \gtrsim -0.015$), and
therefore call for a statistical analysis.

\begin{figure}[t]
\includegraphics[width=0.9\figurewidth]{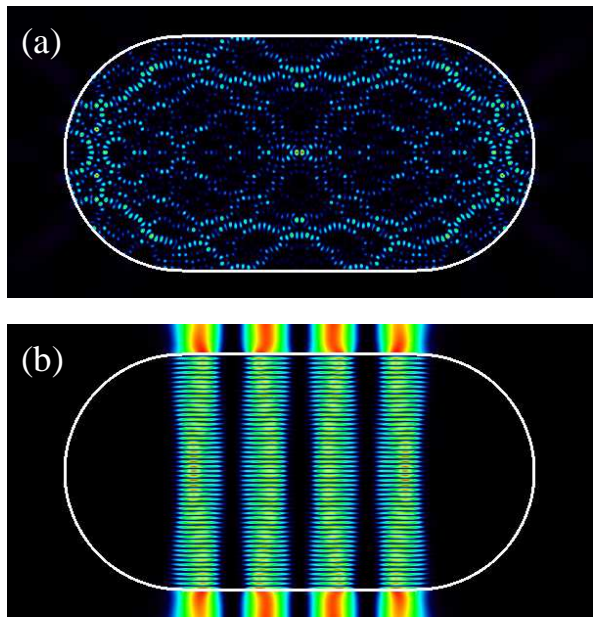}
\caption{(Color online). Near-field pattern of a long-lived (a) and a
  short-lived resonance mode (b) with $\Omega = 15.71886-i0.00337$ and $\Omega =
  15.74567-i0.0378$. As in Fig.~\ref{fig:complexplane}, these modes are computed for TM polarization and
  $n=5$.} \label{fig:lowhighQ}
\end{figure}

Figure~\ref{fig:lowhighQ}(a) and (b) shows examples of long-lived and
short-lived modes in the chaotic microstadium. The complex spatial pattern of
the long-lived mode is in strong contrast to the rather simple pattern of the
short-lived mode. Short-lived modes in the microstadium resemble
bouncing-ball states in the closed stadium billiard \cite{bouncingballs}.
Bouncing-ball modes are usually ignored in the statistical analysis of the
quantum spectrum of closed systems~\cite{Richter}. In the same spirit we
restrict in the following our attention to the long-living resonances, which
are selected via a cut-off of the decay rate ($\im \Omega >-0.04$ for $n=3.3$,
TM; $\im\Omega
>-0.1$ for $n=3.3$, TE; $\im\Omega >-0.015$ for $n=5$, TM; $\im\Omega
>-0.025$ for $n=5$, TE).

\subsection{Determination of characteristic parameters}
A useful reference point for the quantification of the spectral
fluctuations of the long-living resonances are the frequency and time
scales set by their mean spacing along the real frequency axis,
and their mean decay rate. The numerical values of the
dimensionless quantities $(\Delta \Omega)^{(\rm tot)}_{\rm s}$ and
$-\langle\im \Omega \rangle_{\rm s}$ are collected in Tab.\
\ref{tab1}. Here the subscript `s' refers to the stadium billiard,
while the superscript `tot' for the spacing indicates that we do
not discriminate the resonances by parity. We now describe how
this information can be used to determine the parameters $M$ and
$\tau$ which serve as input of the RMT model. Because of parity
conservation, we are interested in the effective parameters per
symmetry class.

\begin{table}[t]
\caption{The first two rows compare the dimensionless mean
frequency spacing of long-living resonances in the stadium
resonator in the interval $66< \re (n\Omega) < 82.5$ (irrespective of
parity) to the prediction from the Weyl law (\ref{eq:weyldelta}).
The second and third row tabulate the mean decay rate of the
long-living resonances and the mean dimensionless decay rate in
the dynamical model (\ref{eq:rmtall}) (obtained in RMT in the
asymptotic limit $M\gg 1$). As described in the text, the relative
scaling of these quantities can be used to determine the time
scale $\tau$ [Eq.\ (\ref{eq:fit1})] and effective matrix dimension
$M$ (for fixed parity) [Eq.\ (\ref{eq:fit2})], which are tabulated
in the last two rows.} \label{tab1}
\begin{tabular}{|l|c|c|c|c|}
\hline \hline
 & $n=3.3$, TM & $n=3.3$, TE & $n=5$, TM & $n=5$, TE
\\
 \hline
$(\Delta \Omega)^{(\rm tot)}_{\rm s}$ &
0.00410 
&
0.00418 
&0.00259  
& 0.00263 
\\
$(\Delta \Omega)^{(\rm tot)}_{\rm Weyl}$ & 0.00359   & 0.00359
&0.00237 & 0.00237
\\        \hline $-\langle\im \Omega \rangle_{\rm s}$ &
0.0143 & 0.0272 &
0.00395
& 0.00736 \\
$-\langle\im \theta \rangle_{\rm RMT}$ & 0.150 &  0.268 & 0.0645&
0.1137 \\ \hline $\tau/\tau_0$ & 1.45 &  1.37  &
1.50
& 1.42 \\
$M$   & 37         & 38  & 37  &  38
\\
\hline\hline
\end{tabular}

\end{table}

Let us first assume that we could ignore the existence of well
organized sequences of short-living resonances, which corresponds
to the case of a fully wave-chaotic resonator. The  mean resonance
spacing then follows the ordinary Weyl law,
\begin{equation}
(\Delta \Omega)^{(\rm tot)}_{\rm Weyl}=\frac{2\pi}{n^2 \re \Omega}
\frac{R^2}{A}, \label{eq:weyldelta}
\end{equation}
where $A=(4+\pi)R^2$ is the area of the billiard. Analogously, the
dimension $M^{(\rm tot)}$ of the scattering matrix can be
estimated by applying the Weyl law to the perimeter $S=(2\pi+4)R$
of the billiard,
\begin{equation}
M^{(\rm tot)}_{\rm Weyl}=\frac{n  \re \Omega}{\pi}\frac{S}{R}.
\label{eq:weylm}
\end{equation}
In the dynamical model, eigenvalue equation (\ref{eq:evals}) has
$M$ solutions with uniform statistical distribution of the phase
$\re \theta$. This translates into a mean dimensionless frequency
spacing
\begin{equation}
(\Delta \Omega)^{(\rm tot)}=\frac{2\pi}{M^{(\rm tot)}}\frac{R}{\tau
c}. \label{eq:weyldyn}
\end{equation}
The combination of Eqs.\ (\ref{eq:weyldelta}), (\ref{eq:weylm}), and
(\ref{eq:weyldyn}) results in the relation
\begin{equation}
\tau=\frac{n \pi A}{c S}
\equiv\tau_0, \label{eq:weyltau}
\end{equation}
which coincides with Sabine's law for the mean propagation time
between encounters with the boundary. If applied to resonances of
a fixed parity, the frequency spacing $\Delta \Omega =4(\Delta \Omega)^{(\rm tot)}$ quadruples, while the effective number of modes
$M=M^{(\rm tot)}/4$ is reduced by a factor four. It follows from
Eq.\ (\ref{eq:weyldyn}) that the time scale $\tau$ is not affected
by these substitutions.

The finiteness of the time to establish wave chaos in an open
system results in the reduction of the number of long-living
quasibound states, which is compensated by the formation of
short-lived states \cite{schomerus:tworzydlo}. A striking example
is the emergence of a fractal Weyl law for systems with a finite
Ehrenfest time
~\cite{fractalweyllaw1,fractalweyllaw2,schomerus:tworzydlo,fractalweyllaw3,fractalweyllaw4}.
 The formation of short-lived states
also often results in  resonance trapping, a phenomenon which
reduces the average decay rate of the long-living resonances~\cite{resonancetrapping}. The existence of
well-organized  bands of resonances deep in the complex-frequency
plane (see Fig.\ \ref{fig:complexplane}) indicates that both
mechanisms are at work in the stadium resonator. As we now show,
the consequences can be described by renormalized parameters $M$
and $\tau$.

Since the reduced number of long-living resonances and resonance
trapping both impact on the average characteristics of complex
frequencies (affecting the real and imaginary part, respectively),
the renormalized parameters can be determined without resorting to
the statistical fluctuations. As a first relation, we compare the
mean decay rate of long-living states in the stadium to the
prediction of RMT,
\begin{equation} \frac{c\tau}{R}=\frac{\langle\im\theta\rangle_{\rm
RMT}}{\langle\im \Omega\rangle_{\rm s}}. \label{eq:fit1}
\end{equation}
For the second relation we evaluate Eq.\ (\ref{eq:weyldyn}) using
the numerically observed frequency spacing of the long-living
resonances, $M^{(\rm tot)}_{\rm s}=[2\pi/(\Delta \Omega)^{(\rm
tot)}_{\rm s}][R/ c\tau]$.
 The
effective number of modes per symmetry class of fixed parity is therefore
given by
\begin{equation} M=\frac{\pi}{2(\Delta \Omega)^{(\rm tot)}_{\rm
s}}\frac{R}{c\tau}.
\label{eq:fit2}
\end{equation}

\begin{figure}[t]
\includegraphics[width=.8\figurewidth]{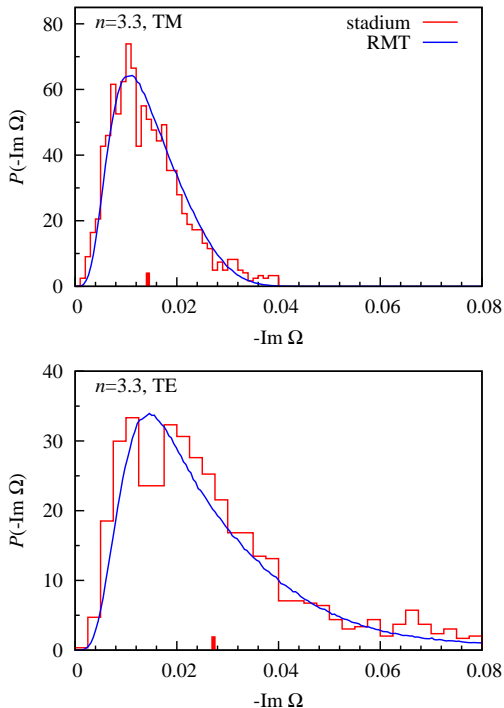}
\caption{(Color online). Distribution of decay rates (in dimensionless units) for
resonances in the stadium resonator with refractive index $n=3.3$, compared to the
prediction of RMT based on the dynamical model (\ref{eq:rmtall})
with $\tau=1.44\tau_0$ and $M=38$.
Short-living resonances in the stadium which form bands in the complex-$\Omega$
plane (see Fig.\ \ref{fig:complexplane}) are neglected via a cut-off. Top panel: TM polarization, $\im\Omega>-0.04$.
Bottom panel: TE polarization, $\im\Omega>-0.1$. The large ticks indicate the value
of the average dimensionless decay rate (see Tab. \ref{tab1}). }
\label{fig:lifetimesn3.3}
\end{figure}

\begin{figure}[t]
\includegraphics[width=.8\figurewidth]{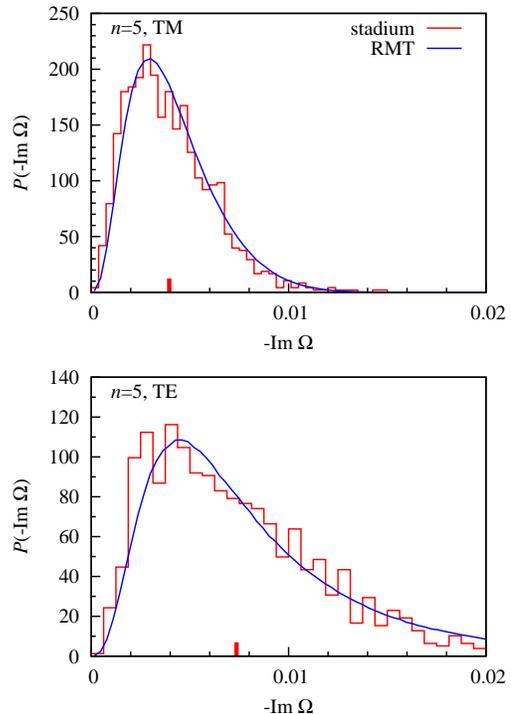}
\caption{(Color online). Same as Fig.\ \ref{fig:lifetimesn3.3}, but for higher
  refractive index, $n=5$. The cut-offs are $\im\Omega>-0.015$ (TM) and $\im\Omega>-0.025$ (TE).
}\label{fig:lifetimesn5}
\end{figure}

In principle, Eqs.\ (\ref{eq:fit1}) and (\ref{eq:fit2}) need to be solved
self-consistently because the RMT average $\langle\im\theta\rangle_{\rm RMT}$
also depends on $M$. For large values of $M$, however, the RMT average
saturates. Practically, therefore, $\tau$ can be obtained from Eq.\
(\ref{eq:fit1}) by fixing $M$ to a large value. The actual value of $M$ then
follows from (\ref{eq:fit2}). This procedure is consistent if the resulting
value $M\gg 1$. Table \ref{tab1} tabulates the values for $\tau$ and $M$
which result when this procedure is applied to the resonances of the stadium
resonator.

For all four combinations of refractive index and polarization, we
find that the collision time $\tau$ differs from the prediction
$\tau_0$ of a totally wave-chaotic resonator by a factor of order
unity, while the effective number of modes $M$ is also almost
identical in all cases. Taking the mean over all four cases we
obtain  the values
\begin{equation}
\tau=1.44\,\tau_0, \quad M=38.
\label{eq:fitvalues}
\end{equation}

\subsection{Life time statistics}
When a resonator is employed as a microlaser, the modes with the
smallest decay rate within the amplification window tend to be the
first that cross the lasing threshold. In a simple model, lasing
occurs as soon as the gain counters the losses through the
interfaces
\cite{ballistic1,ballistic2,ballistic3,ballistic4,random1}. The
analysis of the mean decay rate in the preceding section therefore
delivers a rough estimate of the necessary gain at the lasing
threshold. The distribution of resonances in the complex plane
(Fig.\ \ref{fig:complexplane}), however, suggests that the decay
rates fluctuate broadly. Such large fluctuations entail that some
modes have a much smaller decay rate, which results in a reduced
lasing threshold. We now turn to the question whether these
fluctuations are captured by the RMT variant of the dynamical
model (\ref{eq:rmtall}).

Histograms of the decay rates of the resonances in the stadium
resonators are shown in Figs.\ \ref{fig:lifetimesn3.3} and
\ref{fig:lifetimesn5}. For comparison, the figures also show the
prediction of the RMT model, with parameters given by Eq.\
(\ref{eq:fitvalues}).  Since these parameters apply to the
quasibound states of a fixed parity, the total spectral statistics
are  obtained by the superposition of four systems with identical
effective parameters, but independent microscopic realizations of
the $M$-dimensional random matrix $U$. The RMT results in Figs.\
\ref{fig:lifetimesn3.3} and \ref{fig:lifetimesn5} are based on
50000 realizations per combination of polarization and refractive
index.

In all cases we find excellent agreement of all key features, such
as the modal value of the decay rate, the width of the peak around
the modal value, and the shape of the distribution in the two
tails. In particular, it should be noted that the average decay
rates (tabulated in Tab.\ \ref{tab1}, and shown as large ticks in
the figures) are noticeably larger than the modal value of the
decay-rate distribution. This difference arises from the long tail
of resonances with relatively large decay rate, which closely
follows a slow power law. Such power-law tails are a robust
feature of  RMT models~\cite{fyodorov:sommers},
 including models formed
on the basis of the eigenvalue problem (\ref{eq:evals})~\cite{wei:fyodorov}.
 Furthermore, in the RMT model the width of the
distribution around the modal value depends noticeably on $M$. The
good agreement of the width with the actual histograms lends
direct support to the procedure employed in the determination of
the renormalized effective parameters (\ref{eq:fitvalues}). We
emphasis that the renormalized parameters are determined
independently and do not result from a fitting to the numerically
computed lifetime statistics.

\section{\label{sec:4}Summary}

Dynamical models with scattering matrices of the form
(\ref{eq:rmtall}) have recently attracted considerable attention,
both within RMT~\cite{truncatedrmt1,truncatedrmt2,wei:fyodorov}
 as well as for individual dynamical systems.
In particular, these models have been explored in association with
the Ehrenfest time~\cite{Ehrenfesttimedyn,brouwer,whitney} and the
related question of fractal Weyl
laws~\cite{schomerus:tworzydlo,fractalweyllaw3,fractalweyllaw4,nonnenmacher:schenck}.
Some investigations of the Ehrenfest time employ an effective RMT
model for the scattering matrix
\cite{effectiveRMT1,effectiveRMT2,twofluidmodel}. In these
investigations it was found that certain transport properties
(such as universal conductance fluctuations and shot noise) are
faithfully described by effective RMT, while others (such as the
weak-localization correction) are not \cite{brouwer}.

The present work is motivated by the suggestion in
Ref.~\cite{Keating} that the dynamical model (\ref{eq:rmtall}) can
also be used to describe the spectral fluctuations of long-living
resonances in chaotic dielectric microresonators. We combined this
model with  a variant of effective RMT, which is directly applied
to the internal round-trip operator $U$
\cite{schomerus:tworzydlo}.

We confirm that the RMT variant of the dynamical model captures
the key features of the life time statistics in a specific
dielectric microresonator, the stadium-shaped resonator.  This
comparison requires a careful determination of the effective
collision time $\tau$ of photons with the interfaces and the
effective number of modes $M$ that are mixed by the reflections at
the boundary. Our analysis shows that these parameters can be
reliably obtained by using the mean resonance spacing and decay
rate of the long-living resonances. Compared to the assumption of
complete wave chaos, the effective parameters are renormalized in
a uniform way.
 This renormalization  finds its natural
explanation in the formation of anomalously short-lived resonances
that resemble bouncing-ball modes, which changes the mean
resonance spacing of long-living resonances
\cite{schomerus:tworzydlo}, and the resonance trapping phenomenon
\cite{resonancetrapping}, which changes the mean decay rate.

\begin{acknowledgments}

We would like to thank H.~J. St\"ockmann for discussions.
Financial support from the DFG research group 760 and the European
Commission via the Marie Curie Excellence grant MEXT-2003-02778 is
gratefully  acknowledged.

\end{acknowledgments}


\begin{thebibliography}{28}
\bibitem{remark:correction}  The definition of $p_l$ in Eq.\
(10) of Ref.\ \cite{Keating} contains a misprint. In the continuum
limit $M\to \infty$, $p$ is supposed to cover the range $(-1,1)$.
\end{thebibliography}
\end{document}